\def\rfr#1{eq. (\ref{#1})}

\def\bar{\begin{eqnarray}}
\def\ear{\end{eqnarray}}
\def\bb{\bibitem}
\def\eqi{\begin{equation}}
\def\eqf{\end{equation}}
\def\eqia{\begin{eqnarray}}
\def\eqfa{\end{eqnarray}}
\def\rp#1#2{{#1\over#2}}

\def\lb#1{\label{#1}}
\def\pkp{ P^{\rm Kep} }

\def\oc2{$\mathcal{O}(c^{-2})$}

\documentclass[11pt]{article}
\usepackage{amsmath,amsthm,amscd,amssymb}
\usepackage{latexsym}
\usepackage{graphicx,epsfig}
\usepackage{natbib}
\usepackage{ifpdf}
\usepackage[edtable]{lineno}
\ifpdf
\usepackage[%
  pdftitle={Instructions for use of the document class
    elsart},%
  pdfauthor={Simon Pepping},%
  pdfsubject={The preprint document class elsart},%
  pdfkeywords={instructions for use, elsart, document class},%
  pdfstartview=FitH,%
  bookmarks=true,%
  bookmarksopen=true,%
  breaklinks=true,%
  colorlinks=true,%
  linkcolor=red,anchorcolor=red,%
  citecolor=blue,filecolor=red,%
  menucolor=red,pagecolor=red,%
  urlcolor=cyan]{hyperref}
\else
\usepackage[%
  breaklinks=true,%
  colorlinks=true,%
  linkcolor=red,anchorcolor=red,%
  citecolor=blue,filecolor=red,%
  menucolor=red,pagecolor=red,%
  urlcolor=cyan]{hyperref}
\fi

\begin{document}


\noindent{\bf \LARGE{Determination of tidal distortion in the eclipsing binary system V621 Cen by means of  deviations from the third Kepler law}}
\\
\\
\\
{Lorenzo Iorio}\\
{\it Istituto Nazionale di Fisica Nucleare (INFN), Sezione di Pisa. Viale Unit$\grave{\it a}$ di Italia 68, 70125\\Bari (BA), Italy
\\tel. 0039 328 6128815
\\e-mail: lorenzo.iorio@libero.it}

\begin{abstract}
In this paper   we determine the tidal distortion parameter $k_m$ of the secondary partner (mass loser) of the semi-detached eclipsing binary system V621 Cen by comparing the phenomenologically determined orbital period $P_{\rm b}=3.683549(11)$ to the Keplerian one $\pkp$ computed  with the values of the relevant system's parameters  determined independently of the third Kepler law itself. Our result is $k_m=-1.5\pm 0.6$. Using the periastron precession, as traditionally done with other eclipsing binaries in eccentric orbits, would have not been possible because of the circularity of the V621 Cen path.
\end{abstract}



\section{Introduction}
The dynamics of close binary systems is, in general, different from a purely Keplerian one because of several effects, like tidal and rotational distortions and general relativistic corrections, inducing departures from a simple pointlike two-body Newtonian picture. The measurement of such effects allows to obtain useful information about the physical properties of the system's stars and has been, so far, performed through the detection of the secular precession of the periastron \citep{Cla02} in eccentric systems.  In particular, the tidal and rotational deformation and their impact on the periastron motion were first investigated by \citet{Rus28} and subsequently by \citet{Cow38} and \citet{Ste39}.

In this paper we will consider the Algol-type semi-detached ($R_m/a\sim 0.3$) eclipsing binary system V621 Cen composed by two  stars-the primary, of mass $M$ (mass gainer) and the secondary, of mass $m$, (mass loser) which fills its Roche lobe-in circular orbits with a period\footnote{\citet{Bar07} detected no orbital period's changes during their recent survey.} $P_{\rm b}\approx 3.7$ d. By using its orbital period we will be able to measure the  dynamical tidal distortion parameter $k_m$ of the secondary. V621 Cen has been recently studied both spectroscopically and photometrically by \citet{Bar07}: in Table \ref{tavola} we report the relevant  system's parameters.
\begin{table}
\centering
\begin{minipage}{140mm}
\caption{ Physical and orbital parameters of the V621 Cen binary system estimated by \citet{Bar07} independently of the third Kepler law. The inclination angle $i$ and the stellar masses $M$ and $m$ have been obtained by keeping the temperature $T_M$ fixed to 15,600 K. The relative accuracy is about $1\%$ for all the parameters ($0.03\%$ for the inclination).}
\label{tavola}

\begin{tabular}{@{}lllllll@{}}
\hline
$P_{\rm b}$ (d)& $K_M$ (km s$^{-1}$) & $K_m$ (km s$^{-1}$) & $i$ (deg) & $e$ & $M$ ($M_{\odot}$) & $m$ ($M_{\odot}$) \\
\hline
$3.683549(11)$ & 67.32(1.95) & $197.41(1.59)$ & $76.512(26)$ & $0$ & $6.10(10)$ & $2.09(4)$\\
\hline
\end{tabular}
\end{minipage}
\end{table}
\section{The mass discrepancy}
A first insight that departures from a purely Keplerian picture occur in V621 Cen is the following one.

There is a discrepancy between the sum $\mathcal{M}$ of the masses dynamically determined from the third Kepler law by equating the measured orbital period $P_{\rm b}$ and the computed Keplerian one $\pkp=2\pi\sqrt{a^3/G\mathcal{M}}$, where $a$ is the relative semimajor axis, and the sum of the masses obtained from the non-dynamically inferred values of $M$ and $m$ quoted in Table \ref{tavola}.
To this aim, let us note that such a comparison is, in fact, fully meaningful  because
it is possible to express $a$ and the masses in terms of quantities which have been measured independently of the third Kepler law itself: $K_M$ and $K_m$ from radial velocity spectroscopic analysis and $i, M, m$ (and also $R_M$ and $R_m$) from photometric studies allowed by the eclipsing phenomenon\footnote{$P_{\rm b}$ can be determined both from spectroscopy and photometry, but the latter technique yields more accurate results.}. In general, the projected barycentric semimajor axis of a component of a binary system $x\equiv a_{\rm b}\sin i$ can be obtained from its phenomenologically determined parameters $K$, $P_{\rm b}$ and $e$, if available, as
\eqi x = \rp{KP_{\rm b}}{2\pi}\left(1-e^2\right)^{1/2},\eqf  so that the relative semimajor axis $a$ is
\eqi a = \rp{(K_M+K_m)P_{\rm b}\left(1-e^2\right)^{1/2}}{2\pi\sin i}.\lb{sma}\eqf
%
%

From \rfr{sma} and from the expression of $\pkp$ it follows
\eqi\mathcal{M}_{\rm d}=\left(\rp{P_{\rm b}}{2\pi}\right)\left(\rp{K_M+K_m}{\sin i}\right)^3\left(1-e^2\right)^{3/2}=(7.70\pm 0.22)\mathrm{M}_{\odot}.\lb{somma}\eqf
The uncertainty in \rfr{somma} has been computed by taking the root-sum-square of the individual biased terms due to the uncertainties in $K_m$, $K_M$, $i$ and $P_{\rm b}$; while the errors due to $\delta P_{\rm b}$ and $\delta i$ are of the order of $10^{-5} - 10^{-3}$M$_{\odot}$, $\delta K_M$ and $\delta K_m$ induce bias of the order of $10^{-1}$M$_{\odot}$.  From Table \ref{tavola} it follows
\eqi \mathcal{M}_{\rm nd}=(8.19\pm 0.11)\mathrm{M}_{\odot},\eqf yielding  a $2-\sigma$ discrepancy
\eqi\Delta \mathcal{M}\equiv|\mathcal{M}_{\rm d}-\mathcal{M}_{\rm nd}|=(0.49\pm 0.24)\mathrm{M}_{\odot}.\eqf
Since
\eqi \rp{M}{m}=\rp{K_m}{K_M},\lb{rap_mass}\eqf
we have
\eqi M_{\rm d} = \left(\rp{P_{\rm b}}{2\pi\sin^3 i}\right)\left(1-e^2\right)^{3/2}\left(K_M+K_m\right)^2 K_m = (5.74\pm 0.14){\rm M}_{\odot},\eqf
 \eqi m_{\rm d} = \left(\rp{P_{\rm b}}{2\pi\sin^3 i}\right)\left(1-e^2\right)^{3/2}\left(K_M+K_m\right)^2 K_M = (1.96\pm 0.09){\rm M}_{\odot},\eqf
 with discrepancies
 \eqi \Delta M = (0.36\pm 0.24){\rm M}_{\odot}\ (1.5-\sigma), \eqf
 \eqi \Delta m= (0.13\pm 0.13){\rm M}_{\odot}\ (1-\sigma). \eqf
In order to reconcile $\mathcal{M}_{\rm nd}$ with the third Kepler law, given the measured values of $P_{\rm b}$ and $K_M, K_m$, the inclination should be different from the non-dynamically inferred value quoted in Table \ref{tavola}. Indeed, from \rfr{somma} it turns out that
$i_{\rm d}=72.229\pm 1.879$ deg, which is incompatible at $2.2-\sigma$ level with  $i_{\rm nd}$ by   \citet{Bar07}.
\section{The orbital period and the tidal distortion}
If $\pkp$ is computed with $a$ given by \rfr{sma}   and $\mathcal{M}$ given by $M_{\rm nd} + m_{\rm nd}$ from Table \ref{tavola}
a discrepancy occurs with respect to $P_{\rm b}$.
Indeed\footnote{Here and in the following we will neglect the eccentricity in the formulas since $e=0$ for V621 Cen system.}, \eqi \left(\rp{\pkp}{P_{\rm b}}\right)^2-1= \left(\rp{K_M + K_m}{\sin i}\right)^3\left(\rp{P_{\rm b}}{2\pi G\mathcal{M}_{\rm nd}}\right)-1= -0.06 \pm 0.03,\lb{discr}\eqf
significant at $2-\sigma$ level.
The uncertainty has been worked out by adding in quadrature the following mismodelled terms
\begin{equation}
\left\{
\begin{array}{lll}
3\rp{\left(K_M + K_m\right)^2}{\sin^3 i}\left(\rp{P_{\rm b}}{2\pi G\mathcal{M}_{\rm nd}}\right)\delta K_M = 2.0778\times 10^{-2},\\\\
3\rp{\left(K_M + K_m\right)^2}{\sin^3 i}\left(\rp{P_{\rm b}}{2\pi G\mathcal{M}_{\rm nd}}\right)\delta K_m = 1.6942\times 10^{-2},\\\\
\left(\rp{K_M + K_m}{\sin i}\right)^3\left(\rp{P_{\rm b}}{2\pi G\mathcal{M}^2_{\rm nd}}\right)\delta M_{\rm nd} = 1.1480\times 10^{-2},\\\\
\left(\rp{K_M + K_m}{\sin i}\right)^3\left(\rp{P_{\rm b}}{2\pi G\mathcal{M}^2_{\rm nd}}\right)\delta m_{\rm nd} = 4.592\times 10^{-3},\\\\
3\rp{\left(K_M + K_m\right)^3\cos i}{\sin^4 i}\left(\rp{P_{\rm b}}{2\pi G\mathcal{M}_{\rm nd}}\right)\delta i = 3.07\times 10^{-4},\\\\
\left(\rp{K_M + K_m}{\sin i}\right)^3\left(\rp{P_{\rm b}}{2\pi G^2\mathcal{M}_{\rm nd}}\right)\delta G = 1.41\times 10^{-4},\\\\
\left(\rp{K_M + K_m}{\sin i}\right)^3\left(\rp{1}{2\pi G\mathcal{M}_{\rm nd}}\right)\delta P_{\rm b} = 3\times 10^{-6}.
\end{array}
\right.
\end{equation}
The major sources of uncertainty are represented by $K_M, K_m$ and $M$.

The discrepancy of \rfr{discr} can be interpreted in terms\footnote{In principle, also the general relativistic correction to the third Kepler law \citep{Dam86} is present, but it turns out completely negligible being of the order of $10^{-6}$ d.} of the tidal distortion of $m$. According to the expression for the disturbing radial acceleration for such a case worked out by \citet{Cow38}
\eqi  A_{\rm dist} = -\rp{G(M+m)}{r^2}\left[k_M\rp{ R_M^5}{r^2}\left(\rp{6m}{M r^3} + \rp{\Omega_M^2}{GM}\right)  +
k_m \rp{R_m^5}{r^2}\left(\rp{6M}{m r^3} + \rp{\Omega_m^2}{Gm}\right)\right], \eqf where $\Omega_{m/M}$ are the stellar angular velocities, it turns out \citep{Mar90} that
\eqi \left(\rp{\pkp}{P_{\rm b}}\right)^2-1 = 6k_m\left(\rp{R_m}{a}\right)^5\left(\rp{M}{m}\right),\lb{tidal}\eqf
where $k_m$ is the adimensional tidal coefficient.   By inserting \rfr{sma} into the right-hand-side of \rfr{tidal} and expressing the ratio of the masses as in \rfr{rap_mass}, \rfr{tidal}, \rfr{discr} and $R_m=5.87(1)$R$_{\odot}$ \citep{Bar07} yield
\eqi k_m = -1.5\pm 0.6,\lb{kappa}\eqf
accurate at $2.5-\sigma$ level.  The major sources of error are $K_m$, $K_M$, $M$ and $m$ which yield bias of the order of $10^{-1}$; $\delta R_m$ affects $k_m$  at a $10^{-2}$ level, $\delta G$ and $\delta i$ at $10^{-3}$, while the bias due to $\delta P$ is $10^{-5}$.
We have neglected both the centrifugal terms, proportional to the square of the stellar angular velocities, and the tidal term of $M$ which is smaller than the one of $m$ by one order of magnitude.


The fact that $k_m$ is negative and larger than unity is in agreement with the theory of dynamical tide \citep{Wil04} and with the determined values for many eccentric binary systems from their periastron precessions \citep{Cla02}.

It maybe interesting to note that for the (detached, $R/a\sim 0.1$) eclipsing binary system with circular orbit  WW Aurig{\ae}  \citep{Sou05} the discrepancy between $\pkp$, computed as here for V621 Cen, and $P_{\rm b}$ is well compatible with zero being $(\pkp/P_{\rm b})^2-1 = (0.09\pm 5.18)\times 10^{-3}$; the tidal components of the orbital period are of the order of $10^{-4}$ d for both stars, while the uncertainty in the Keplerian one is about one order of magnitude larger.
The same holds for the (semi-detached, $R_m/a = 0.25$)  eclipsing binary  UNSW-V-500 \citep{scuti} for which  $(\pkp/P_{\rm b})^2-1 = (0.3\pm 5)\times 10^{-2}$; in this case   the tidal distortion of $m$ affects the orbital period at $10^{-2}$ d level, while the bias due to the Keplerian component amounts to $10^{-1}$ d.

\section{Conclusions}
In this paper we showed that in the eclipsing semi-detached binary  system V621 Cen, whose physical and orbital parameters were recently determined from accurate photometric/spectroscopic studies, a discrepancy significant at $2-\sigma$ level occurs between the sum of the masses $\mathcal{M}_{\rm nd}$ dynamically determined from the third Kepler law and the one inferred non-dynamically. By interpreting such a deviation from the third Kepler law as due to the tidal distortion of the secondary partner (mass loser) of mass $m$ we were able to determine its effective Love number getting $k_m = -1.5\pm 0.6$. Such a result points towards the existence of dynamical tide effects in V621 Cen.  It is important to note that, being the orbit of V621 Cen circular, the periastron rate is not available, contrary to other eccentric eclipsing binaries for which $k$ was determined with such a method. Our result, significant at $2.5-\sigma$ level, should encourage further studies of V621 Cen.

\section*{Acknowledgments}
I gratefully thank F. Barblan for useful discussions.


\end{document}